\documentclass[12pt]{article}
\usepackage{amssymb}
\def\bea{\begin{eqnarray}}
\def\eea{\end{eqnarray}}
\def\be{\begin{equation}}
\def\ee{\end{equation}}
\def\re#1{(\ref{#1})}

\begin{document}

\begin{titlepage}
\vskip 1. cm
\begin{center}
\huge{\bf{Kaluza-Klein towers for spinors in flat space}}
\end{center}
\vskip 0.6 cm
\begin{center} {\Large{Fernand Grard$^1$}},
\ \ {\Large{Jean Nuyts$^2$}}
\end{center}
\vskip 1 cm

\noindent{\bf Abstract}
\vskip 0.2 cm
{\small
\noindent
Considering a massive or massless free spinor field propagating 
in a flat five dimensional space with its fifth dimension compactified
either on a strip or on a circle, we analyse the procedure of 
generation of the four dimensional Kaluza-Klein spinor mass towers.
Requiring the five dimensional Dirac operator to be symmetric, 
the set of all the allowed boundary conditions is obtained. 
In the determination of the boundary conditions
and in the Kaluza-Klein reduction equations, 
the SO(3,1) and parity invariances in the space-time subspace 
are carefully taken into account. The equations determining
the mass towers are written in full generality. 
A few numerical examples are given.    

 }
\vfill
\noindent
{\it $^1$  Fernand.Grard@umh.ac.be,  Physique G\'en\'erale et
Physique des Particules El\'ementaires,
Universit\'e de Mons-Hainaut, 20 Place du Parc, 7000 Mons, Belgium}
\vskip 0.2 cm
\noindent
{\it $^2$  Jean.Nuyts@umh.ac.be,
Physique Th\'eorique et Math\'ematique,
Universit\'e de Mons-Hainaut,
20 Place du Parc, 7000 Mons, Belgium}
\end{titlepage}

\section{Introduction \label{Introd}}

In recent articles,
we have reanalysed in a mathematically complete and fully consistent
way, the generation of 
Kaluza-Klein mass towers \cite{KK} in five dimensional theories 
with a compactified fifth dimension. This study was
carried out for a scalar field 
supposed to propagate in the bulk, at first in
a flat space \cite{GN1}
and then in a warped space without metric singularities
\cite{GN2} and in a warped space with metric singularities \cite{GN3}.
The mathematical approach relies heavily
on a precise study of the 
hermiticity properties of the Kaluza-Klein reduction equations which are
of second order in derivatives.

This has allowed us to classify all the sets of allowed boundary
conditions.  These sets include, as particular cases, 
the usual box and periodic or antiperiodic
boundary conditions which are currently invoked.
We found, as a main result, that the Kaluza-Klein mass states may form
non-regular towers which depend 
on the specific set of boundary conditions considered
for the various possible metric configurations. 

As the future high energy colliders
will look for the possible
appearance of Kaluza-Klein mass towers 
as evidence for the existence of fields propagating in
higher dimensions and that these towers, if they exist, may
well be composed of spinor states, 
we were led to extend our work to the
Dirac fields. At first sight, this problem appeared simpler as the Dirac 
equation is of first order only in derivatives. However,
the presence of multicomponents spinors offers new perspectives 
and hence 
increases the complexity of the solutions.

In this article, we restrict ourselves, 
as a first step in a more general approach, 
to a five-dimensional flat space. 
This leads to a convenient toy model
where the degrees
of freedom of a free Dirac field propagating in the bulk 
already show up and play a major role.

The article is organized as follows. In Section \re{secDirac}, 
we recall a few properties of the Dirac equation in five dimensions, 
putting forward the specific characteristics that are needed to 
construct and classify the Kaluza-Klein towers.
In Section \re{secKKreduc}, taking into account
the underlying invariances and symmetries,
in particular covariance and parity invariance in the 
four dimensional space time, the 
Kaluza-Klein reduction equations together with the set 
of all allowed boundary conditions are established.
The resulting mass equations, from which the Kaluza-Klein mass towers are
built, are given in Section \re{secKKmass}.
In Section \re{secnum}, some numerical examples are presented and discussed. 

Our approach 
will be extended 
in a forthcoming article
to a 
five dimensional warped space, 
which is known 
to provide a natural and elegant solution to the hierarchy problem \cite{RS}
as an alternative to the solution based on large extra dimensions 
and gravity considerations
\cite{ADD}. Its 
peculiar
characteristics will allow, in particular, 
the generation of Kaluza-Klein towers
with an expected more realistic physical content. 
 
\section{Dirac equation in a five-dimensional flat space \label{secDirac}}

We consider a free spinor field with mass $M$ 
satisfying the five dimensional Dirac equation
\be
\left(i\gamma^a\partial_a -M\right)\Psi=0 \ .
\label{Diracflat5}
\ee
The field is
supposed to propagate in the bulk,
a flat five dimensional space with coordinates 
$x^a, \ a\equiv \{\mu,5\}\equiv \{\mu,s\},\ \mu=0,1,2,3$ 
and a metric
\begin{equation}
{\rm{diag}}(\eta_{_{_{ab}}})
     =\{+1,-1,-1,-1,-1\}
\label{etaflat5}
\end{equation}
giving rise to an invariance group SO(4,1) 
(related for the spinor representation to the symplectic group Sp(4)). The fifth dimension 
$s$ is compactified 
either on a strip or on a circle ($0\leq s \leq 2\pi R$).
 
The five Dirac matrices in this space satisfy
\begin{equation}
\left[\gamma^{a},\gamma^{b}\right]_{+}
      =2\eta^{ab}\ .
\label{gaeqflat5}
\end{equation}
There are two inequivalent sets of 
$\gamma^a$ matrices which can be built from the four usual 4$\times$4 Dirac matrices
$\gamma^{\mu}$ and
\be
\gamma^{5}=\gamma^{0}\gamma^{1}\gamma^{2}\gamma^{3}\quad,\quad\left(\gamma^5\right)^2=-1\ .
\label{gamma5flat}
\ee 
They are  
\begin{equation}
\gamma^{[I]a}\equiv \{\gamma^{\mu},\gamma^5\}
\label{gammaIflat5}
\end{equation}
or
\begin{equation}
\gamma^{[II]a}\equiv \{-\gamma^{\mu},-\gamma^5\}\ .
\label{gammaIIflat5}
\end{equation}
Contrary to the four dimensional case, there is no transformation
mapping one set to the other 
\be 
\left\{\ \exists \hspace{-2.8 mm} \diagdown\ P \mid\ \gamma^{[II]a}=P\gamma^{[I]a}P^{-1}\right\}\ .
\label{minusgamflat5}
\ee
If there is another set of $4\times 4$ $\gamma^{a}$ matrices
satisfying \re{gaeqflat5}, this set is equivalent through a change of basis,
either
to the set $\gamma^{[I]a}$ or to the set $\gamma^{[II]a}$. 
In particular, the sets $\left(\gamma^a\right)^{+}$ 
and $\left(\gamma^a\right)^{t}$ satisfy \re{gaeqflat5} and are equivalent
to the set $\gamma^{[I]a}$
\bea
\left(\gamma^a\right)^{+}&=&A\gamma^aA^{-1}\quad,\quad A=\gamma^0
      \label{transflat5}\\
\left(\gamma^a\right)^{t}&=&D\gamma^aD^{-1}\quad,\quad D=-D^t=C\gamma^{5}     
\label{adjflat5}
\eea
where $C$ is the usual four dimensional 
charge conjugation matrix satisfying
$C\gamma^{\mu}C^{-1}=-\left(\gamma^{\mu}\right)^t$ 
and where the antisymmetric matrix $D$ is related 
to the symplectic metric of Sp(4).

It should be remarked, taking into account \re{minusgamflat5}, that $M$
and $-M$ correspond to distinct fields.
The covariance, under the covering group of SO(4,1), 
of the Dirac equation 
in a five dimensional space follows exactly the same pattern 
as the four dimensional one. In particular, the infinitesimal 
generators of the spinor transformations $\psi'(x')=S\psi(x)$
are $\sigma^{ab}=\frac{i}{4}[\gamma^a,\gamma^b]_{_-}$ 
and are identical for the two $\gamma$ representations
\re{gammaIflat5}, \re{gammaIIflat5}. 

For spinor fields, one uses
the natural invariant hermitian scalar product 
between two spinors $\phi$ and $\psi$
(with as usual $\overline{\phi}=\phi^{+}\gamma^0$)
\begin{equation}
(\phi,\psi)_{\rm{hermitian}}=\int_{x=-\infty}^{+\infty}\int_{s=0}^{2\pi R}\,
{\overline{\phi}}\,\psi\, d^4 x\, ds 
\label{scalhermiflat5}
\end{equation}
and not the invariant symplectic scalar product
\begin{equation}
(\phi,\psi)_{\rm{symplectic}}=\int_{x=-\infty}^{+\infty}\int_{s=0}^{2\pi R}\,
\phi^t\,D\,\psi\, d^4 x\, ds \ .
\label{scalsymplecflat5}
\end{equation}

The Dirac operator ${\cal{D}}=i\gamma^a\partial_a$
is symmetric for all $\phi,\psi$ in its domain
\begin{equation}
\left(\phi,{\cal{D}}\phi\right)_{\rm{hermitian}}=\left({\cal{D}}\phi,\phi\right)_{\rm{hermitian}}\ ,
\label{symmetryflat5} 
\end{equation}
provided the following
boundary relation is satisfied
\begin{equation}
\Biggl[\int_{-\infty}^{+\infty}{\overline{\phi}}\gamma^5\psi\,d^4x\Biggr]_{s=2\pi R}=
     \Biggl[\int_{-\infty}^{+\infty}{\overline{\phi}}\gamma^5\psi\,d^4x\Biggr]_{s=0}\ .
\label{BCeqintflat5}
\end{equation}
In Appendix \re{AppendixBC}, we justify with heuristic arguments 
the more restrictive condition which we will impose
\begin{equation}
\Bigl[{\overline{\phi}}\gamma^5\psi\Bigr]_{s=2\pi R}=
     \Bigl[{\overline{\phi}}\gamma^5\psi\Bigr]_{s=0}
\label{BCeqflat5}
\end{equation}
for all values of $x^{\mu}$.
This implies 
the existence of at least four linear equations 
among the components of the fields evaluated at  
$s=2\pi R$ and $s=0$. 
These boundary conditions should respect the SO(3,1) invariance
in the four dimensional subpace $x^{\mu}$.
Hence we postulate for the boundary conditions 
the general form
$\bigl[\psi\bigr]_{s=2\pi R}
= \left(c_1 \,1\hspace{-1.45 mm}1_4+c_2\gamma^5\right)\bigl[\psi\bigr]_{s=0}$
with two complex constants $c_1$ and $c_2$ 
($1\hspace{-1.45 mm}1_4$ is the unit matrix in spinor space).
Introducing this form in the restriction \re{BCeqflat5}, one finds
that the coefficients $c_1$ and $c_2$ are expressible in terms of a real
parameter $\omega$ with infinite extend 
and a phase angle $-\pi\leq \rho <\pi$
\bea
\biggl[\psi\biggr]_{s=2\pi R}= e^{i\rho}\left(\cosh(\omega) 1\hspace{-1.45 mm}1_4
                        +i \sinh(\omega)\gamma^5\right)\biggl[\psi\biggr]_{s=0}\ .
\label{BC2flat5}
\eea 
This is the natural set of boundary conditions in $s$ 
valid for all $x^{\mu}$ 
within the hypothesis of SO(3,1) covariance. As will be seen later, 
this form of the boundary conditions 
does not imply violation of parity in any four-dimensional brane. 
This is due to the fact that the $\gamma^{5}\partial_5$ part 
in the Dirac equation induces 
a subtle natural connection between $\psi(x^{\mu},s)$ and
its derivative multiplied by $\gamma^{5}$. 

\section{Kaluza-Klein reduction for the Dirac equation \label{secKKreduc}}
 
The   
Kaluza-Klein reduction of the spinor $\psi_{\alpha}(x^{\mu},s)$ is carried out
assuming the separation of variables
\begin{equation}
\psi_{\alpha}(x^{\mu},s)
    =\sum_n \biggl(F^{[n]}(s)1\hspace{-1.45 mm}1_4
                 +i G^{[n]}(s)\gamma^{5}\biggr)\psi_{\alpha}^{[n]}(x^{\mu})
\label{KKreducflat5}
\end{equation}
with, for each $n$,
two complex scalar functions of $s$, $F^{[n]}(s)$ and $G^{[n]}(s)$,
and a spinor $\psi_{\alpha}^{[n]}(x^{\mu})$ depending on $x^{\mu}$. 
This form requires SO(3,1) covariance only which allows the presence of the $\gamma^{5}$ term.
The $i$ has been put for convenience.

Introducing this Kaluza-Klein ansatz in the Dirac equation 
\re{Diracflat5}, one obtains
\bea
    &   \biggl(F^{[n]}1\hspace{-1.45 mm}1
            -iG^{[n]}\gamma^{5}\biggr)
       \biggl(i\gamma^{\mu}\partial_{\mu}\psi_{\alpha}^{[n]}(x^{\mu})\biggr)&
    \nonumber\\
    &\ -\biggl(\left(M F^{[n]} -\partial_s G^{[n]}\right)1\hspace{-1.45 mm}1
                +i\left(M G^{[n]} - \partial_s F^{[n]} \right)\gamma^5
       \biggr)       
       \biggl(
       \psi_{\alpha}^{[n]}(x^{\mu})
       \biggr)
       =0\ ,&    
\label{Diracreducflat5}
\eea
while the boundary conditions \re{BC2flat5} become
\be
\left(\matrix{F^{[n]}(2\pi R)\cr
              G^{[n]}(2\pi R) }\right)
=e^{i\rho}\left(
            \matrix{
            \cosh(\omega)&\sinh(\omega)\cr
            \sinh(\omega)&\cosh(\omega) }\right)
           \left(\matrix{F^{[n]}(0)\cr
                         G^{[n]}(0)}\right) \ .
\label{BCreducflat5}
\ee
One sees that the boundary conditions imply that 
the values of $F^{[n]}$ and $G^{[n]}$ at $2\pi R$ and at $0$
must be related by a U(1)$\times$SO(1,1) transformation.
Remark that the usual periodic or antiperiodic
boundary conditions 
for the spinors (allowing to closure 
of the $s$ strip to a circle) correspond 
to the case $\omega=0$ and $\rho=0$ or $\rho=\pi$ 
\be
F^{[0]}(2\pi R)=\epsilon\, F^{[0]}(0)
\quad,\quad G^{[0]}(2\pi R)=\epsilon\, G^{[0]}(0)\quad {\rm{with\ }}
\epsilon^2=1\ .
\label{periodBCflat5}
\ee

In a four-dimensional physical brane, 
we request that the spinor $\psi_{\alpha}^{[n]}(x^{\mu})$
should satisfy the parity invariant
Dirac equation
\begin{equation}
i\gamma^{\mu}\partial_{\mu}\psi_{\alpha}^{[n]}(x^{\mu})
    =m_n \psi_{\alpha}^{[n]}(x^{\mu})
\label{Diracbraneflat5}
\end{equation}
with, by convention, $m_n\geq 0$.
Then, $\psi_{\alpha}(x^{\mu},s)$ from \re{KKreducflat5} 
is a solution of the Dirac equation in five dimensions if 
the two following coupled equations for $G^{[n]}(s)$ and $F^{[n]}(s)$
are satisfied
\bea
\partial_s G^{[n]}(s)&=&(M-m_n)F^{[n]}(s)
    \nonumber\\
\partial_s F^{[n]}(s)&=&(M+m_n)G^{[n]}(s)  \ .
\label{Diracreduc2flat5}
\eea
With the boundary conditions \re{BCreducflat5} taken into account, 
these equations lead to the determination of the allowed 
spinor masses $m_n$ observable in a four dimensional subspace
\re{Diracbraneflat5}. 

\section{Kaluza-Klein mass towers \label{secKKmass}}

In this section, we treat successively 
the distinct cases corresponding to the bulk mass $M$ being
positive, zero or negative. 

\subsection{The case $M>0$ \label{subsecMpositive}}

Remember that $m_n$ is positive by convention.

\subsubsection{The subcase $m_n^2>M^2$ ($M>0$) \label{subsecmM}}
For $m_n^2>M^2$, the solutions of \re{Diracreduc2flat5} are
\bea
F^{[n]}(s)=\sqrt{m_n+M}\left(\sigma_n \sin\left(\sqrt{m_n^2-M^2} s\right)
                              +\tau_n\cos\left(\sqrt{m_n^2-M^2} s\right)\right)&&
    \nonumber\\                      
&&\label{solFG1flat5}
             \\                       
G^{[n]}(s)=\sqrt{m_n-M}\left(\sigma_n \cos\left(\sqrt{m_n^2-M^2} s\right)
                          -\tau_n\sin\left(\sqrt{m_n^2-M^2} s\right)\right)&&
   \nonumber
\eea
where $\sigma_n$ and $\tau_n$ are constants and the square roots are chosen positive.

Introducing these solutions in the set of 
boundary conditions \re{BCreducflat5}, 
one obtains a system of two linear homogeneous equations for $\sigma_n$ 
and $\tau_n$. The vanishing of the related determinant 
gives the mass equation for the $m_n$'s 
\bea
&&\left(\cosh(\omega)\cos\left(2\pi \sqrt{m_n^2-M^2}\, R\right) 
-\cos\left(\rho\right)\right)\sqrt{m_n^2-M^2}
      \nonumber\\
&&\quad\quad\quad= M\sinh(\omega) \sin\left(2\pi\sqrt{m_n^2-M^2}\, R\right) \ .  
\label{tower4flat5}
\eea
Note the scaling property of the equation, that it does not depend on $R$ when
the masses are expressed in units of $1/R$
\bea 
m_n &=& \frac{\overline{m}_n}{R}
        \nonumber\\
M &=&\frac{\overline{M}}{R} \ .
\label{defscalflat5}
\eea
In general, for given values of $\omega$, $\rho$ and $\overline{M}$, 
this equation has an infinite number of solutions $\overline{m}_n$, 
giving rise to a Kaluza-Klein tower.

Asymptotically, for large $n$, i.e. when $m_n>> M$, the masses in the tower are given by
\be
\cos\left(2\pi \overline{m}_n\right)\approx\frac{\cos(\rho)}{\cosh(\omega)}
\quad \left(\overline{m}_n>>\overline{M}\right)\ .
\label{asymptoticflat5}
\ee 
and become identical to the masses 
in the $M=0$ tower for the same boundary parameters $\rho$ and
$\omega$ \re{Mzero}.

\subsubsection{The subcase $m_1=M$ ($M>0$) \label{subsecm1M}}

For  $m_1 = M$, the solutions are
\bea
G^{[M]}(s)&=&\sigma_M
    \nonumber\\
F^{[M]}(s)&=&2M\sigma_M s+\tau_M
\label{solFG2flat5}
\eea
where $\sigma_M$ and $\tau_M$ are constants.
Introducing  
these solutions 
in the boundary conditions \re{BCreducflat5}, 
one finds two linear homogeneous
relations in the parameters $\sigma_M$ and $\tau_M$. 
Defining $\overline{M}_1$ as
\begin{equation}
\overline{M}_1=\frac{\cosh(\omega)-\cos(\rho)}{2\pi \sinh(\omega) }\ ,
\label{MRunflat5}
\end{equation}
the vanishing of the determinant leads to the following condition
\be
\overline{M}=\overline{M}_1 
\label{condmMflat5}
\ee
to be satisfied by the parameters $\omega$, $\rho$ and $\overline{M}$
for the first mass in the tower $m_1$ 
to be equal to the bulk mass  
\be
m_1=\,\vert M \vert\ .
\label{mMflat5}
\ee

\subsubsection{The subcase $m_{h}^2<M^2$ ($M>0$) \label{subsechM}}

For $m_{h}^2 < M^2$, the solutions are
\bea
F^{[h]}(s)\hspace{-3 mm}&=&\hspace{-3 mm}\sqrt {M+m_{h}}\left(\sigma_h \sinh\left(\sqrt{M^2-m_{h}^2} s\right)
                                +\tau_h\cosh\left(\sqrt{M^2-m_{h}^2} s\right)\right)
    \nonumber\\
    &&\label{solFG4flat5}\\
G^{[h]}(s)\hspace{-3 mm}&=&\hspace{-3 mm}\sqrt{M-m_{h}}\left(\sigma_h \cosh\left(\sqrt{M^2-m_{h}^2} s\right)
                                +\tau_h\sinh\left(\sqrt{M^2-m_{h}^2} s\right)\right)\ .
                                \nonumber
\eea
Replacing these solutions into the boundary conditions,
one again finds two linear homogeneous equations in $\sigma_h$ and $\tau_h$.
The determinant is zero provided
\bea
&\left(\cosh(\omega)\cosh\left(2\pi \sqrt{M^2-m_{h}^2}\,R\right)-\cos(\rho)\right)\sqrt{M^2-m_{h}^2}&
      \nonumber\\
&\quad\quad      
    =M\sinh(\omega)\sinh\left(2\pi \sqrt{M^2-m_{h}^2}\,R\right)\ .&
\label{hyperb4flat5}
\eea
For given values of 
the boundary parameters $\omega$, $\rho$ and of $\overline{M}$ \re{defscalflat5},
the solution of this equation, if it exists, is unique and will be the
lowest mass $m_1=m_h$ in the tower, such that
\be
0<m_{1}<\vert M\vert\ .
\label{hypermassflat5}
\ee
The formula \re{hyperb4flat5} is simply 
the analytical continuation of  \re{tower4flat5}.

\subsubsection{The subcase $m_{h}=0$ ($M>0$) \label{subsech0}}

For $m_{h}=0$, the limiting case of the equation \re{hyperb4flat5}, namely
\be
\cosh\left(2\pi M R-\omega\right)=\cos(\rho) \ ,    
\label{hyperm0flat5}
\ee
implies, with the definition
\be
\overline{M}_2=\frac{\omega}{ 2\pi}\ ,
\label{MRdeuxflat5}
\ee
the following restrictions
on the parameters $\omega$, $\rho$ and $\overline{M}$ 
\be
\rho=0\quad,\quad \overline{M}=\overline{M}_2
\label{m0equal0flat5}
\ee
for a zero mass state to exist.

\subsubsection{Summary \label{summary}}

The results (for $M>0$) related to the presence or absence of a first mass in the tower
lower than the bulk mass $M$ are summarized in Appendix \re{appendixM}. 

\subsection{The case $M=0$    \label{Mzero}}

The case of the bulk mass $M=0$ is obtained by letting $M\rightarrow 0$ 
in the relevant formulas.

\subsubsection{The subcase $m_n>0$ ($M=0$) \label{subsecm0}}

For $m_n>0$, 
the solutions of \re{Diracreduc2flat5} are
\bea
F^{[n]}(s)&=&\sigma_n \sin\left(m_n s\right)
                              +\tau_n\cos\left(m_n s\right)
    \nonumber\\                      
G^{[n]}(s)&=&\sigma_n \cos\left(m_n s\right)
                          -\tau_n\sin\left(m_n s\right)
\label{solFG1M0flat5}
\eea
where $\sigma_n$ and $\tau_n$ are constants.
After introducing these solutions in the boundary conditions \re{BCreducflat5}, 
the vanishing of the related determinant 
gives the mass equation for the $m_n$ Kaluza-Klein tower 
\be
\cosh(\omega)\cos\left(2\pi m_n R\right) 
-\cos\left(\rho\right)=0    \ .  
\label{tower4M0flat5}
\ee
The $\overline{m}_n$ tower is the superposition of two regular subtowers, each with 
spacing 
\be
\Delta(n+2,n)\equiv\overline{m}_{n+2}-\overline{m}_{n}=1\ .
\label{spacing1flat5}
\ee 
The separation between the two subtowers is given by
\be
\Delta(2n+1,2n)\equiv
\overline{m}_{2n+1}-\overline{m}_{2n}=\frac{1}{\pi}\arccos\left(\frac{\cos(\rho)}{\cosh(\omega)}\right)
\ .
\label{spacing2flat5}
\ee

\subsubsection{The subcase $m_1=0$ ($M=0$) \label{subsecmM0}}

To obtain a zero mass state ($m_1=0$), the lowest in the tower, 
one sees from \re{Diracreduc2flat5} that $F^{[n]}$ and $G^{[n]}$ must be constants  
and hence, from \re{BCreducflat5},
that the boundary conditions must be $\omega=\rho=0$. 
This corresponds to the periodic
boundary conditions \re{periodBCflat5} with $\epsilon=1$,
allowing the closure of the $s$ strip to a circle.

\subsection{The case $M<0$    \label{Mnegative}}

The case $M<0$ is analogous to the case $M>0$. 
As the main result, the mass towers are related as follows 
\be
{\rm{tower\,}}\biggl\{-M,-\omega,\rho\biggr\}\ \equiv\  {\rm{tower\, }}\biggl\{M,\omega,\rho\biggr\}\ .
\label{minusMflat5}
\ee

\section{Numerical evaluations \label{secnum}}

Illustrative numerical examples of spinor towers
are presented in the five tables for a representative set of
bulk masses $\overline{M}$ and for some chosen values 
of the boundary parameters $\omega$ and $\rho$.

As a general comment, for $\overline{M}=0$, 
there are two interlaced regular subtowers, 
each with equal spacing $\Delta(n+2,n)=1$ \re{spacing1flat5}
and variable separation $\Delta(2n+1,2n)$ \re{spacing2flat5} between  
the odd and even indexed masses.

\begin{enumerate}

\item 

In Table \re{tablemoins1pisur3}, the boundary parameters are
$\omega=-1$ and $\rho=\pi/3$. 
Since $\omega$ is negative,
the masses appearing in the tower are always larger 
than the bulk mass $M$, as it should
for any value of $\rho$. 
For $M=0$, the even-odd separation is 
\be
\Delta(2n+1,2n)\approx 0.395\ .
\label{separation1flat5}
\ee       
When the bulk mass $\overline{M}$ increases, the first masses, say the eight first masses $\overline{m}_1,\dots,\overline{m}_8$, 
become closer and closer to $\overline{M}$. 
Already at $\overline{M}=100$, one sees that these first masses become very densely packed
just above $\overline{M}$. 
However, in all cases, asymptotically in $n$, 
the mass towers all tend to the mass tower corresponding to $M=0$.

\item

In Table \re{table0pisur3}, the boundary parameters are
$\omega=0$ and $\rho=\pi/3$. 
For $M=0$, the even-odd separation is 
\be
\Delta(2n+1,2n)\approx 0.333\ .
\label{separation2flat5}
\ee       
For increasing $\overline{M}$, the towers behave as in Table \re{tablemoins1pisur3}.

\item

In Table \re{table0dixieme}, the boundary parameters are
$\omega=0$ and $\rho=0.1$. Compared to the table \re{table0pisur3}, one sees that,
for $M=0$, 
the even-odd separation has become smaller 
\be
\Delta(2n+1,2n)\approx 0.032\ .
\label{separationf3flat5}
\ee       
Indeed, at the limit of $\rho=0$, corresponding to the periodic boundary conditions
\re{periodBCflat5},
the two subtowers merge for $M=0$. The mass $\overline{m}_1$ is zero while the other masses
($\overline{m}_n=n,\, n\neq 0$) are doubly degenerate. 
For increasing $\overline{M}$ and asymptotically, the towers behave as before.

\item

In Table \re{table2pisur3}, the boundary parameters are
$\omega=2$ and $\rho=\pi/3$. 
For $M=0$, the even-odd separation is 
\be
\Delta(2n+1,2n)\approx 0.458\ .
\label{separationf4flat5}
\ee   
For 
$\overline{M}<\overline{M}_1$ \re{MRunflat5}, $\overline{m}_1$ is larger than $\overline{M}$. 
For $\overline{M}>\overline{M}_1$, it is smaller than $\overline{M}$.  
Disregarding the exceptional $\overline{m}_1$, 
the towers behave as before
for increasing $\overline{M}$ and asymptotically.

\item

In Table \re{table20}, the boundary parameters are
$\omega=2$ and $\rho=0$. 
For $M=0$, the even-odd separation is 
\be
\Delta(2n+1,2n)\approx 0.414\ .
\label{separationf5flat5}
\ee    
For 
$\overline{M}<\overline{M}_1$, $\overline{m}_1$ \re{MRunflat5} is larger than $\overline{M}$. 
For $\overline{M}>\overline{M}_1$, it is smaller than  $\overline{M}$. 
For $\overline{M}=\overline{M}_2$
\re{MRdeuxflat5}, there is a zero mass state in the tower
($\overline{m}_1=0$). 
Here again, disregarding the exceptional $\overline{m}_1$, 
the towers behave as before
for increasing $\overline{M}$ and asymptotically.

\end{enumerate}

\section{Conclusions \label{Conclu}}

In this article, we have carefully analysed the procedure of generation of Kaluza-Klein 
mass towers of four dimensional spinor fields, starting 
from a massive or massless five dimensional free Dirac field 
propagating 
in a flat bulk space with its fifth dimension compactified 
on a strip or on a circle.

Requiring the five dimensional Dirac operator to be symmetric,
we have deduced
the set of all the allowed boundary conditions.
The natural 
invariant hermitian scalar product and the
SO(3,1) invariance in the $x^{\mu}$ subspace were taken into account. 
The boundary conditions depend in a subtle way on the properties of the 
$\gamma^{5}$ matrix and are expressible in terms of two free parameters.

The Kaluza-Klein reduction is conducted in such a way that the spinor fields
in four dimensions, which are related to a given bulk spinor field,
obey the ordinary parity invariant Dirac equation. 
Requiring SO(3,1) covariance, it turns out that the $\gamma^{5}$ matrix 
plays also an essential role in the separation of variables.  
Notwithstanding, the presence of $\gamma^{5}$ does not spoil the parity conservation.

The equations whose solutions provide the Kaluza-Klein mass towers 
have been written in full generality. A few numerical 
examples are presented and discussed.

This work will be extended to the expected more realistic case of
spinor fields propagating in five dimensional 
warped spaces, in line with
our recent model of scalar fields 
living in warped spaces without \cite{GN2} 
or with \cite{GN3} metric singularities.

\vspace{5cm}

\noindent{\bf{Acknowledgments}}

The authors would like to thank Dr. Gregory Moreau
for suggesting that, in our research program, the fermion case
was to be given high priority.

\newpage

\appendix

\section{Heuristic justification of \re{BC2flat5} \label{AppendixBC}}

In this appendix, we justify with plausibility 
and simplicity arguments our 
derivation of the form \re{BC2flat5} for the general boundary conditions
for the Dirac fields.

The integrated boundary condition \re{BCeqintflat5}
\begin{equation}
\Biggl[\int_{-\infty}^{+\infty}{\overline{\phi}}\gamma^5\psi\,d^4x\Biggr]_{s=2\pi R}=
     \Biggl[\int_{-\infty}^{+\infty}{\overline{\phi}}\gamma^5\psi\,d^4x\Biggr]_{s=0}
\label{ap1flat5}
\end{equation} 
should lead to linear relations between the fields 
$\psi(x^{\mu},s)$ evaluated at the edges of the $s$ domain, namely at $s=2\pi R$ and $s=0$.
Given $\psi(y^{\mu},0)$ for all $y^{\mu}$, 
$\psi(x^{\mu},2\pi R)$ would then be related to it by the most general linear
relation
\be
\psi_{\alpha}(x^{\mu},2\pi R)
=\int_{-\infty}^{+\infty} C_{\alpha\beta}(x^{\mu},y^{\nu})\,\psi_{\beta}(y^{\nu},0)\,d^4y
\label{linearelflat5}
\ee
where $C_{\alpha\beta}(x^{\mu},y^{\nu})$ is a complex 4$\times$4 matrix 
of functions depending on the eight coordinates.

If this boundary condition is to be 
covariant under the space-time SO(3,1) (subgroup of SO(4,1)) transformations 
$\psi'(x^{'\mu},s)_{\alpha}=S_{\alpha\beta}\,\psi_{\beta}(x^{\mu},s)$,
the matrix $C_{\alpha\beta}$ must 
in particular commute with $S$ and hence is restricted 
to a combination of the unit and $\gamma^5$ matrices
\be
C(x^{\mu},y^{\nu})=C_1(x^{\mu},y^{\nu}) 1\hspace{-1.45 mm}1_4 
                 +iC_2(x^{\mu},y^{\nu})\gamma^5
\label{Ciflat5}                 
\ee
where $C_1(x^{\mu},y^{\nu})$ and $C_2(x^{\mu},y^{\nu})$ are 
two complex invariant functions or distributions which depend essentially
on the invariant distance $(x-y)^2$ between the points $x^{\mu}$ and $y^{\mu}$
(the $i$ is for convenience).

Introducing this form \re{Ciflat5}, \re{linearelflat5}, 
valid both for $\psi$ and for $\phi$, 
in the condition \re{ap1flat5}, one finds
\bea
\int_{-\infty}^{+\infty} \biggl(C_1^*(x,y)C_1(x,z)-C_2^*(x,y)C_2(x,z)\biggr)d^4x&=&\delta^4(y-z)
      \nonumber\\
\int_{-\infty}^{+\infty} \biggl(C_1^*(x,y)C_2(x,z)-C_2^*(x,y)C_1(x,z)\biggr)d^4x&=&0\ .
\label{apeq2flat5}
\eea
The natural solution is expressible in terms of invariant $\delta$ distributions
\be
C_i(x^{\mu},y^{\nu})=c_i\,\delta^4(x^{\mu}-y^{\mu})
\label{apeq3flat5}
\ee
with the two constants 
being $c_1=e^{i\rho}\sinh(\omega)$ and 
$c_2=e^{i\rho}\cosh(\omega)$, leading 
through \re{linearelflat5} to 
the final form \re{BC2flat5}. More general solutions of \re{apeq2flat5}
are probably not very useful.

\section{Summary of the results for $M>0$}{\label{appendixM}}

Let us summarize the results for the position 
of the lowest mass $m_1$ in a tower relative to the bulk mass $M>0$.

\begin{enumerate}

\item

For $\omega<0$, there is no mass $m_{h}< M$ (see \re{hyperb4flat5})
and hence $m_1>M$.

\item

For $\omega=0$ and $\rho\neq 0$, there is no mass $m_{h}< M$.
Indeed, the condition resulting from \re{hyperb4flat5}
($\cosh(2\pi \sqrt{\overline{M}^2-\overline{m}_{h}^2})=\cos(\rho)$) 
has no solution. Hence $m_1>M$.

\item

For $\omega=0$ and $\rho=0$, the lowest mass in the tower is $m_{1} =M_{M}=M$. 

\item

For $\omega>0$ and $\overline{M}<\overline{M}_3$, 
\bea
\overline{M}_3&=&\frac{\cosh(\omega)-1}{2\pi \sinh(\omega)}
           \label{MRtroisflat5}\\
\overline{M}_4&=&\frac{\cosh(\omega)+1}{2\pi \sinh(\omega)}\ ,
            \label{MRquatflat5}
\eea
there is no mass $m_{h} < M$.  Hence $m_1>M$.

\item

For $\omega>0$, 
$\overline{M}_3\leq\overline{M}\leq\overline{M}_4$ and $-\rho_1<\rho<\rho_1 $, 
\be
\rho_{1}=\arccos\biggl(\cosh(\omega)-2\pi \sinh(\omega)\overline{M}\biggr) 
\quad\quad  \left(0<\rho_1<\pi\right)\ ,
\label{rho1flat5}
\ee
there is a mass $m_{h}< M$,
which is the lowest mass ($m_1=m_{h}$) in the tower.
For $\rho=\rho_1$, $m_1=M=M_1$ \re{MRunflat5}.

\item

For $\omega>0$, 
$\overline{M}>\overline{M}_4$, there is a mass $m_{h}<M$ for any value of $\rho$. 
The lowest mas $m_1$ is always smaller than $M$.

\item

A mass $m_0=0$ exists provided that the boundary conditions belong to the case
\be
\rho=0\quad ,\quad \omega=2\pi \overline{M}\ . 
\ee
This mass is the lowest mass in the tower.
\end{enumerate}

\newpage
\vskip 0.5 cm

\begin{table}
\caption{
Mass towers for $\omega=-1$ and $\rho=\frac{\pi}{3}$
{\label{tablemoins1pisur3}}
}
\vspace{1 cm}
\hspace{1 cm}
\scriptsize{
\begin{tabular}{|c|c|c|c|c|c|c|c|c|}
\hline
\multicolumn{9}{|c|}
      {$\phantom{\bigl[\bigr.}$ Case $\omega=-1$, $\rho=
      \frac{\pi}{3}$}
         \\
\multicolumn{9}{|c|}
      {$\phantom{\bigl[\bigr.}$For very large $n$, the mass
      towers converges toward the $\overline{M}=0$ tower }
        \\ \hline
  $\phantom{\bigl[\bigr.}\overline{M}$
        &$\overline{m}_1$&$\overline{m}_2$
        &$\overline{m}_3$&$\overline{m}_4$
        &$\overline{m}_5$&$\overline{m}_6$
        &$\overline{m}_7$&$\overline{m}_8$
      \\ \hline
0  &  0.198&0.803&1.198&1.803&2.198&2.803&3.198&3.803
      \\ \hline
0.1&  0.267&0.823&1.212&1.812&2.205&2.809&3.203&3.807
      \\ \hline
0.2&  0.346&0.854&1.234&1.827&2.218&2.818&3.211&3.814
      \\ \hline
0.3&  0.431&0.895&1.264&1.847&2.234&2.831&3.223&3.824
      \\ \hline
0.4&  0.519&0.943&1.301&1.871&2.256&2.848&3.238&3.836
      \\ \hline
0.5&  0.609&0.998&1.344&1.901&2.281&2.867&3.255&3.851
      \\ \hline
0.6&  0.700&1.059&1.392&1.935&2.310&2.890&3.276&3.868
      \\ \hline
0.7&  0.793&1.126&1.446&1.974&2.343&2.917&3.300&3.888
      \\ \hline
1  &  1.077&1.346&1.632&2.112&2.464&3.013&3.387&3.961
      \\ \hline
2  &  2.049&2.210&2.413&2.758&3.048&3.502&3.835&4.346
      \\ \hline
10 &10.012&10.0486&10.106&10.193&10.292&10.429&10.566
&10.751
      \\ \hline
100&100.001&100.005&100.011&100.02&100.031&100.045&100.061
&100.08
      \\ \hline
\end{tabular}
   }
\end{table}

\vskip 0.5 cm

\begin{table}
\caption{
Mass towers for $\omega=0$ and $\frac{\pi}{3}$
{\label{table0pisur3}}
}
\vspace{1 cm}
\hspace{1 cm}
\scriptsize{
\begin{tabular}{|c|c|c|c|c|c|c|c|c|}
\hline
\multicolumn{9}{|c|}
      {$\phantom{\bigl[\bigr.}$ Case $\omega=0$, $\rho=
      \frac{\pi}{3}$}
         \\
\multicolumn{9}{|c|}
      {$\phantom{\bigl[\bigr.}$For very large $n$, the mass
      towers converge toward the $\overline{M}=0$ tower }
        \\ \hline
  $\phantom{\bigl[\bigr.}\overline{M}$
      &$\overline{m}_1$&$\overline{m}_2$
      &$\overline{m}_3$&$\overline{m}_4$
      &$\overline{m}_5$&$\overline{m}_6$
      &$\overline{m}_7$&$\overline{m}_8$
      \\ \hline
0  &0.167&0.833&1.167&1.833&2.167&2.833&3.167&3.833
      \\ \hline
0.1&0.194&0.839&1.171&1.836&2.169&2.835&3.168&3.835
      \\ \hline
0.2&0.260&0.857&1.184&1.844&2.176&2.840&3.173&3.837
      \\ \hline
0.3&0.343&0.886&1.205&1.858&2.187&2.849&3.181&3.845
      \\ \hline
0.4&0.433&0.924&1.233&1.877&2.203&2.862&3.192&3.854
      \\ \hline
0.5&0.527&0.972&1.269&1.900&2.224&2.877&3.206&3.866
      \\ \hline
0.6&0.623&1.027&1.312&1.929&2.248&2.896&3.223&3.880
      \\ \hline
0.7&0.720&1.088&1.361&1.963&2.277&2.919&3.243&3.897
      \\ \hline
1  & 1.014&1.302&1.537&2.088&2.386&3.005&3.321&3.962
      \\ \hline
10 &10.001&10.035&10.068&10.167&10.232&10.394&10.489&10.710
      \\ \hline
100&100.0001&100.0034&100.0068&100.0168&100.0235&100.0401
&100.0501&100.0734
      \\ \hline
\end{tabular}
   }
\end{table}

\begin{table}
\caption{
Mass towers for $\omega=0$ and $\rho=0.1$
{\label{table0dixieme}}
}
\vspace{1 cm}
\hspace{1 cm}
\scriptsize{
\begin{tabular}{|c|c|c|c|c|c|c|c|c|}
\hline
\multicolumn{9}{|c|}
      {Case $\phantom{\bigl[\bigr.}\omega=0$, $\rho=
      0.1$ }
         \\
\multicolumn{9}{|c|}
      {$\phantom{\bigl[\bigr.}$ For very large $n$, the mass
      towers converge to the $M=0$ tower }
        \\ \hline
  $\phantom{\bigl[\bigr.}\overline{M}$
        &$\overline{m}_1$&$\overline{m}_2$
        &$\overline{m}_3$&$\overline{m}_4$
        &$\overline{m}_5$&$\overline{m}_6$
        &$\overline{m}_7$&$\overline{m}_8$
      \\ \hline
0  &  
0.016&0.984&1.016&1.984&2.016&2.984&3.016&3.984
      \\ \hline
0.1&
0.101&0.989&1.021&1.987&2.018&2.986&3.018&3.985
      \\ \hline
0.2&  
0.201&1.004&1.035&1.994&2.026&2.991&3.023&3.989     
      \\ \hline
0.3&
0.3004&1.029&1.059&2.007&2.038&2.999&3.031&3.995
      \\ \hline
0.4&
0.4003&1.062&1.092&2.024&2.055&3.011&3.042&4.004
      \\ \hline
0.5&
0.5003&1.104&1.132&2.046&2.077&3.026&3.052&4.015
      \\ \hline
0.6&
0.6002&1.153&1.180&2.073&2.103&3.044&3.075&4.029
      \\ \hline
0.7&
0.7002&1.208&1.234&2.104&2.134&3.065&3.096&4.045
      \\ \hline
1  &
1.0001&1.403&1.426&2.222&2.250&3.147&3.177&4.108
      \\ \hline
2  &
2.00006&2.229&2.243&2.817&2.840&3.592&3.619&4.458
      \\ \hline
5  &
5.000025&5.096&5.102&5.379&5.391&5.823&5.839&6.393
      \\ \hline
10 &
10.000013&10.048&10.051&10.195&10.201&10.436&10.445&10.764
      \\ \hline
100&
100.0000013&100.0048&100.0052&100.0197
&100.0203&100.0445&100.0455&100.0793
      \\ \hline
\end{tabular}
   }
\end{table}

\vskip 0.5 cm
\begin{table}
\caption{
Mass towers for  $\omega =2$ and $\rho=\frac{\pi}{3}$
{\label{table2pisur3}}
}
\vspace{1 cm}
\hspace{1 cm}
\scriptsize
{
\begin{tabular}{|c|c|c|c|c|c|c|c|c|c|c|}
\hline
\multicolumn{10}{|c|}
      {$\phantom{\bigl[\bigr.}$ Case $\omega=2$, $\rho=
      \frac{\pi}{3}$}
        \\
\multicolumn{10}{|c|}
      {$\phantom{\bigl[\bigr.}$
       $\overline{M}_1=0.143\dots$  \re{MRunflat5},
       $\overline{M}_3=0.121\dots$  \re{MRtroisflat5},
       $\overline{M}_4=0.209\dots$ \re{MRquatflat5}
      }
         \\
\multicolumn{10}{|c|}
      {$\phantom{\bigl[\bigr.}$ For very large $n$, the mass
      towers converge to the $M=0$ tower }
        \\ \hline
$\phantom{\bigl[\bigr.}\overline{M}$ &$\rho_1$\re{rho1flat5}
     &$\overline{m}_1$&$\overline{m}_2$&$\overline{m}_3$
     &$\overline{m}_4$&$\overline{m}_5$&$\overline{m}_6$
     &$\overline{m}_7$&$\overline{m}_8$
      \\ \hline
0&&
0.229&0.771&1.229&1.771&2.229&2.771&3.229&3.771
      \\
0.1&&
0.167&0.757&1.220&1.765&2.224&2.768&3.226&3.769
      \\
0.12 &&
0.155&0.756&1.220&1.765&2.224&2.767&3.225&3.768
      \\
$\overline{M}_3$&0&
0.155&0.756&1.220&1.765&2.224&2.767&3.225&3.768
      \\
0.13 &0.644&
0.150&0.756&1.220&1.765&2.224&2.767&3.225&3.768
      \\
0.14 &0.962&
0.145&0.755&1.220&1.765&2.224&2.767&3.225&3.768
      \\
$\overline{M}_1$&$\rho$&
$\overline{M}_1$
&0.755&1.219&1.765&2.224&2.767&3.225&3.768    
      \\
0.15&1.220&
0.140&0.755&1.219&1.764&2.224&2.767&3.225&3.768        
      \\
0.20&2.490&
0.117&0.756&1.220&1.765&2.224&2.767&3.226&3.768       
      \\
$\overline{M}_4$&$\pi$&
0.114&0.757&1.220&1.765&2.224&2.768&3.226&3.768    
      \\
0.5&&
0.118&0.834&1.271&1.798&2.251&2.788&3.244&3.784     
      \\
1&&
0.264&1.166&1.502&1.960&2.384&2.894&3.336&3.862    
      \\
2&&
0.532&2.074&2.268&2.577&2.912&3.334&3.727&4.199    
      \\
10&&
2.658&10.013&10.051&10.116&10.203&10.319&10.453
&10.616                                            
      \\
100&&
26.585&100.0012&100.005&100.011&100.020&100.031
&100.045&100.061                                    
      \\ \hline
\end{tabular}
   }
\end{table}

\vskip 0.5 cm
\begin{table}
\caption{
Mass towers for $\omega=2$ and $\rho=0$
{\label{table20}}
}
\vspace{1 cm}
\hspace{1 cm}
\scriptsize
{
\begin{tabular}{|c|c|c|c|c|c|c|c|c|c|}
\hline
\multicolumn{9}{|c|}
      {$\phantom{\bigl[\bigr.}$ Case $\omega=2\quad$, $\quad\rho=0$}
        \\ 
\multicolumn{9}{|c|}
      {$\phantom{\bigl[\bigr.}$
       $\overline{M}_1=\overline{M}_3=0.121\dots$ \re{MRunflat5}, \re{MRtroisflat5}, 
       $\overline{M}_2=1/\pi=0.318\dots$         \re{MRdeuxflat5} 
       }
         \\
\multicolumn{9}{|c|}
      {$\phantom{\bigl[\bigr.}$ For very large $n$, the mass
      towers converge to the $M=0$ tower }       
        \\ \hline
$\phantom{\bigl[\bigr.}\overline{M}$ &$\overline{m}_1$&$\overline{m}_2$&$\overline{m}_3$
                 &$\overline{m}_4$&$\overline{m}_5$&$\overline{m}_6$
                 &$\overline{m}_7$&$\overline{m}_8$
      \\ \hline
0       &0.207   &0.793  &1.207  &1.793  &2.207  &2.793  &3.207  &3.793 
      \\
0.05    &0.171   &0.785  &1.202  &1.789  &2.204  &2.791  &3.205  &3.791 
      \\
0.1     &0.136   &0.779  &1.199  &1.787  &2.203  &2.789  &3.204  &3.790 
      \\
$\overline{M}_1$(=$\overline{M}_3)$&$\overline{M}_1$&0.778  &1.198  &1.786  &2.202  &2.789  &3.204  &3.790 
      \\
0.15    &0.102   &0.777  &1.198  &1.786  &2.202  &2.789  &3.204  &3.790
      \\    
0.2    &0.069    &0.778  &1.199  &1.787  &2.202  &2.789  &3.204  &3.790 
      \\
$\overline{M}_2$&0       &0.793  &1.211  &1.793  &2.208  &2.793  &3.208  &3.793 
      \\
0.4    &0.043    &0.813  &1.225  &1.802  &2.216  &2.799  &3.213  &3.798 
      \\
1      &0.262    &1.173  &1.489  &1.977  &2.366  &2.913  &3.316  &3.882 
      \\
10     &2.658    &10.013 &10.052 &10.117 &10.203 &10.321 &10.451 &10.619  
      \\
100    &26.580   &100.001&100.005&100.011&100.020&100.031&10.045 &100.061  
      \\ \hline
\end{tabular}
   }
\end{table}

\end{document}